\documentclass[%
superscriptaddress,
reprint,
nofootinbib,
nobibnotes,
aps,
prl,
floatfix]{revtex4-1}

\usepackage{lmodern}
\usepackage{amsmath,bm}
\usepackage{amssymb,amsfonts}
\usepackage{xcolor}
\usepackage{graphicx}
\usepackage{bm}
\usepackage{enumitem}
\usepackage{mathtools}
\usepackage[colorlinks=true,linkcolor=blue,citecolor=blue,urlcolor=blue]{hyperref}
\usepackage{cleveref}

\providecommand{\xhat}{}\renewcommand{\xhat}{\hat{\bm{x}}}
\providecommand{\yhat}{}\renewcommand{\yhat}{\hat{\bm{y}}}
\providecommand{\zhat}{}\renewcommand{\zhat}{\hat{\bm{z}}}
\providecommand{\nhat}{}\renewcommand{\nhat}{\hat{\bm{n}}}
\providecommand{\Rhat}{}\renewcommand{\Rhat}{\hat{\bm{R}}}

\providecommand{\vmu}{}\renewcommand{\vmu}{\bm{\mu}}
\providecommand{\vr}{}\renewcommand{\vr}{\bm{r}}
\providecommand{\vB}{}\renewcommand{\vB}{\bm{B}}
\providecommand{\vd}{}\renewcommand{\vd}{\bm{d}}
\providecommand{\vR}{}\renewcommand{\vR}{\bm{R}}
\providecommand{\Z}{}\renewcommand{\Z}{\mathbb{Z}}

\graphicspath{ {./images/} }
\begin{document}

\title{Magnetic Dipole in a Cuboidal Superconducting Trap}

\author{Francis J. Headley}
\affiliation{Institut f\"ur Theoretische Physik,
Eberhard-Karls-Universit\"at T\"ubingen, 72076 T\"ubingen, Germany}

\date{\today}

\begin{abstract}
We derive the exact image-dipole potential of a point dipole inside a closed cuboidal superconducting trap. The construction generalises the parallel-plate result to a geometry that confines every translational degree of freedom, and we prove that the image lattice satisfies the Meissner boundary condition on all six walls. For a centred dipole the orientational energy reduces to a diagonal quadratic form whose three coefficients are Epstein-zeta-type lattice sums. We show that in both the infinite and finite rectangular traps the dipole orientation aligns with the \emph{short} cross-sectional axis over a finite range of aspect ratios. The equilibrium orientation in both cases is described by a phase diagram whose degeneracies we classify. Every prediction is verified against finite-element solutions of the same boundary-value problem, with agreement better than $0.16\%$.
\end{abstract}

\maketitle

\section{Introduction}

Magnetic levitation, famously popularised by Sir Michael Berry and André Geim~\footnote{Much to the dismay of amphibians everywhere.} \cite{BerryFrog}, has transitioned from being a physics novelty to being a powerful tool in the field of quantum sensing \cite{gonzalez-ballestero_levitodynamics_2021}. Levitated magnetic systems, currently proposed and utilised in dark matter searches~\cite{AmaralDM}, precision metrology~\cite{Fuchs2024, HeadleyNewton}, quantum tunnelling~\cite{HeadleyRot}, and tests of new forces~\cite{AmaralNonNewton}, present themselves as an exciting low-energy platform from which we can probe the laws of fundamental physics.

Among these platforms, magnets levitated in superconducting traps stand at the forefront, offering exceptional isolation from environmental noise \cite{gonzalez-ballestero_levitodynamics_2021} and the prospect of long mechanical coherence times \cite{Fuchs2024,timberlake_acceleration_2019,Vinante2020}. The stability and dynamics of such levitated magnets have been studied in both theory~\cite{Lin2006,RusconiLinear,RusconiRigid} and experiment~\cite{prat2017,Vinante2020,Uitenbroek2026}. The image method was applied in Ref.~\cite{HeadleyMagnet} to a point dipole situated between two infinite parallel superconducting plates, yielding the trapping potential in closed form. In this geometry the dipole is confined both in its position along the axis normal to the plates and in its librational tilt relative to them. Due to translational symmetry of the system, the dipole remains free to drift parallel to the plates unhindered.

Experimentally, however, these superconducting traps are often fully enclosed, rather than a pair of infinite plates, and it is this full three-dimensional confinement that both localises the particle and selects its equilibrium orientation. The present work treats a magnetic dipole in a closed rectangular trap, for which the trap is confining in every translational coordinate. We begin by establishing that the image construction is consistent: reflecting the source in one pair of walls must not violate the boundary condition on the additional orthogonal walls. Then, as the additional confinement introduces an orientational degree of freedom whose equilibrium is set by the trap shape, we reduce the orientational energy to a diagonal quadratic form, before showing that the resulting easy-axis selection is surprisingly non-monotonic in the trap aspect ratio. Every prediction is verified against a finite-element solution of the same boundary-value problem. 

\section{Image construction}
A superconductor in the Meissner state is a perfect diamagnet: the magnetic field $\vB$ vanishes in its interior of the superconductor, and continuity of the normal field component across the boundary $\partial\Omega$ of the trap region $\Omega$ imposes the boundary condition
\begin{equation}
\label{eq:BC}
   \vB(\vr)\cdot\nhat=0,\qquad \vr\in\partial\Omega,
\end{equation}
where $\nhat$ is the surface normal. The tangential field is unconstrained; it is supported by screening currents within a London-penetration layer, which is negligibly thin on the millimetre scale of interest. We solve the resulting Neumann problem by the method of images, building up from a single plane to the closed trap.

For a single infinite superconducting plane, the boundary condition \eqref{eq:BC} is enforced by adding one image dipole on the far side of the plane~\cite{vinante_levitated_2022,HeadleyMagnet}. The plane acts as a mirror: the image sits at the mirror-image position, with the moment component normal to the plane reversed and the tangential components preserved. For a plane normal to $\zhat$ the moment transforms as
\begin{equation}
\label{eq:single}
   (\mu_x,\mu_y,\mu_z)\longmapsto(\mu_x,\mu_y,-\mu_z).
\end{equation}
We call the image produced by this rule the \emph{mirror} of the dipole in that wall. Between two parallel plates a single image is not sufficient: the mirror image dipole that cancels the normal field on one of the plates in turn produces a normal field contribution on the other plate. We must therefore create a new image dipole on the opposide side of said plate, and repeating this generates an infinite chain of images of alternating orientation, whose sum is the unique solution of Laplace's equation with the prescribed dipole source~\cite{HeadleyMagnet}.

\begin{figure}[t]
  \centering
  \includegraphics[width=\columnwidth]{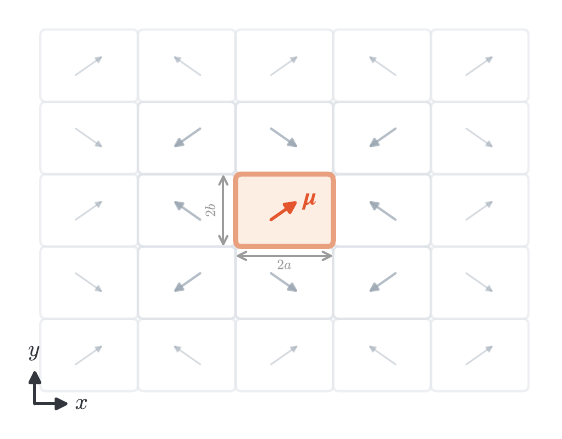}
  \caption{Image construction in the cross-section of a rectangular tube. The real dipole $\vmu$ (red) sits at the centre of the trap $\Omega$ of half-widths $a$ and $b$. Mirroring the dipole in a wall places an image dipole at the mirror position and reverses the moment component normal to that wall, generating the lattice \eqref{eq:lattice}. The moment direction therefore alternates with the parities of the cell indices $(n_x,n_y)$.}
  \label{fig:lattice}
\end{figure}

To begin, we now place the dipole in a rectangular trap of domain
\begin{equation}
\label{eq:cavity}
   \Omega=[-a,a]\times[-b,b]\times[-c,c],
\end{equation}
with half-widths $a$, $b$ and $c$ along $\xhat$, $\yhat$ and $\zhat$, respectively, with the point dipole being of moment $\vmu=(\mu_x,\mu_y,\mu_z)$, magnitude $\mu=|\vmu|$, and held at $\vr_0\in\Omega$. The trap has six walls, and we mirror the dipole in each of them. Mirroring in two orthogonal walls commutes, since each changes only the coordinate and the moment component normal to its own plane, so the order in which the walls are used does not matter. Mirroring the dipole, and in turn all of the image dipoles, repeatedly in all six walls therefore generates a unique lattice of images, indexed independently along the three axes,
\begin{equation}
\label{eq:lattice}
\begin{aligned}
   \vr_{\bm n}&=\bigl(2n_xa+(-1)^{n_x}x_0,\ 2n_yb+(-1)^{n_y}y_0,\\
   &\hphantom{{}=\bigl(}\qquad\qquad \qquad\qquad \ \ 2n_zc+(-1)^{n_z}z_0\bigr),\\[3pt]
   \vmu_{\bm n}&=\bigl((-1)^{n_x}\mu_x,\,(-1)^{n_y}\mu_y,\,(-1)^{n_z}\mu_z\bigr),
\end{aligned}
\end{equation}
where the integer triple $\bm n=(n_x,n_y,n_z)\in\Z^3$ indexes the infinite image dipoles across the $x$-, $y$- and $z$-walls, with $\bm n=\bm 0$ being the original dipole. The two-dimensional version of \eqref{eq:lattice} is shown in Fig.~\ref{fig:lattice}: the images tile the plane, and the moment alternates direction from cell to cell with the parities of $n_x$ and $n_y$. By construction the lattice is closed under mirroring in every wall, so each image dipole has a mirror partner across each of the six walls. As shown in the next section, it is this pairing that enforces the boundary condition \eqref{eq:BC} on all six walls at once.

\section{Boundary condition}
\label{sec:bc}

As said before, an image introduced to correct the boundary condition on one wall contributes a normal field on all of the others, so it is not immediate that the completed lattice \eqref{eq:lattice} satisfies \eqref{eq:BC} on all six walls. The proof of the cancellation rests on a single identity, established below by direct calculation, together with the reflection symmetry of the lattice. 

\paragraph{Pair identity.}
Let two dipoles be related by reflection in the plane $z=c$: a dipole $\vmu$ at $\vr_s=(x_s,y_s,z_s)$ and its image $\vmu^\ast=(\mu_x,\mu_y,-\mu_z)$ at $\vr_s^\ast=(x_s,y_s,2c-z_s)$. Their combined field has vanishing normal component on the plane $z=c$, as we now verify. Let $\vr=(x,y,c)$ be any point of the plane and write $\vd=\vr-\vr_s$. Reflection sends $z_s\mapsto2c-z_s$, so that the displacement from the image is $\vd^\ast=\vr-\vr_s^\ast=(d_x,d_y,-d_z)$, which has the same magnitude $R=|\vd|=|\vd^\ast|$ as $\vd$. The normal component of a dipole field is
\begin{equation}
   B_z=\frac{\mu_0}{4\pi R^3}\Bigl[\tfrac{3(\vmu\cdot\vd)\,d_z}{R^2}-\mu_z\Bigr].
\end{equation}
For the image, $\vmu^\ast\cdot\vd^\ast=\mu_xd_x+\mu_yd_y+(-\mu_z)(-d_z) =\vmu\cdot\vd$, the normal displacement is $d_z^\ast=-d_z$, and the normal moment is $\mu_z^\ast=-\mu_z$, such that
\begin{equation}
\label{eq:pair}
   B_z^{\ast}
   =\frac{\mu_0}{4\pi R^3}\Bigl[\tfrac{3(\vmu\cdot\vd)(-d_z)}{R^2}+\mu_z\Bigr]
   =-\,B_z .
\end{equation}
The two normal fields cancel at every point of the plane. The identity holds for any pair related by reflection in $z=c$ and the corresponding moment map, independently of the side of the plane on which the dipoles lie.

\paragraph{Reflection symmetry of the lattice.}
Let $\sigma$ denote reflection in $z=c$, acting on positions by $(x,y,z)\mapsto(x,y,2c-z)$. Applying this to \eqref{eq:lattice} leaves the $x$ and $y$ entries unchanged and sends the $z$ coordinate $z_{n_z}=2n_zc+(-1)^{n_z}z_0$ to
\begin{equation}
   2c-z_{n_z}=2(1-n_z)c+(-1)^{1-n_z}z_0=z_{1-n_z},
\end{equation}
using $(-1)^{1-n_z}=-(-1)^{n_z}$. Hence $\sigma$ permutes the lattice sites by the index map $(n_x,n_y,n_z)\mapsto(n_x,n_y,1-n_z)$. The moment carried by the image site is
\begin{equation}
\begin{aligned}
   \vmu_{(n_x,n_y,1-n_z)}&=\bigl((-1)^{n_x}\mu_x,(-1)^{n_y}\mu_y,
       (-1)^{1-n_z}\mu_z\bigr)\\
   &=\bigl((-1)^{n_x}\mu_x,(-1)^{n_y}\mu_y,-(-1)^{n_z}\mu_z\bigr),
\end{aligned}
\end{equation}
which is exactly the reflected moment of $\vmu_{\bm n}$ under \eqref{eq:single}. Therefore $\sigma$ maps the entire set of dipoles onto itself and sends each moment to its image, such that each site and its partner $\bm n\leftrightarrow(n_x,n_y,1-n_z)$ satisfy \eqref{eq:pair}.

\paragraph{Convergence.}
Let $\rho$ be the distance from a wall point to an image $\bm n$; the images sit on the lattice \eqref{eq:lattice}, and a single image contributes a normal field of order $\rho^{-3}$. In the rectangular tube the lattice is two-dimensional: a shell $[\rho,\rho+\mathrm{d}\rho]$ holds $O(\rho)\,\mathrm{d}\rho$ images, the summed magnitudes are bounded by $\int\!\mathrm{d}\rho\,\rho\cdot\rho^{-3}<\infty$, and the field sum converges absolutely. In the closed trap the lattice is three-dimensional: a shell holds $O(\rho^{2})\,\mathrm{d}\rho$ images, the same bound becomes $\int\!\mathrm{d}\rho\,\rho^{-1}$, and the magnitudes sum to a logarithmic divergence. The signed sum still converges, through the angular cancellation of the dipole field, but only \emph{conditionally}, so its value can depend on the order of summation.

The summation order is nonetheless fixed and the shape dependence of a conditionally convergent dipole sum enters only through the demagnetising field of the net magnetisation of the summation region. The image moments alternate from cell to cell, so the lattice has zero net magnetisation and no such term arises. Every region that expands symmetrically about the source therefore yields the same limit, which we adopt.

\paragraph{Cancellation on the wall.}
The index map $n_z\mapsto1-n_z$ has no fixed point, since $n_z=1-n_z$ has no integer solution; equivalently no image lies on the wall, because $z_{n_z}=c$ has no solution for an interior source, $|z_0|<c$. Evaluating the shape-independent limit over a region symmetric under $\sigma$, the images partition into disjoint mirror pairs $\{\bm n,(n_x,n_y,1-n_z)\}$, and by the pair identity \eqref{eq:pair} each pair contributes zero normal field at every point of $z=c$; the boundary condition \eqref{eq:BC} follows. The same argument applies to the other five walls through the involutions $n_z\mapsto-1-n_z$, $n_x\mapsto\pm1-n_x$ and $n_y\mapsto\pm1-n_y$, so a single lattice satisfies all six boundary conditions at once.

\paragraph{Uniqueness.}
The regularised field is divergence- and curl-free in $\Omega$ with the prescribed dipole singularity and, by the argument above, satisfies \eqref{eq:BC} on all six walls; the interior Neumann problem has exactly one such solution, and any additive uniform field would violate \eqref{eq:BC} on at least one wall pair. The same regularised limit makes the trap energy well defined. The finite-element solution of Fig.~\ref{fig:fem} confirms that the regularised field satisfies the boundary condition to within $0.16\%$.

\section{Trap potential and orientational energy}

The field induced at the source is the sum of the fields of the image dipoles, $\vB_{\rm ind}(\vr_0)$. As the screening currents that produce this field are themselves induced by the source, the trapping potential carries the standard image self-energy factor of one half, in the same way as for an electrostatic image charge~\cite{Jackson1998,vinante_levitated_2022,Lin2006},
\begin{equation}
   U(\vr_0,\vmu)=-\tfrac12\,\vmu\cdot\vB_{\rm ind}(\vr_0).
\end{equation}
Written as an explicit sum:
\begin{equation}
\label{eq:U3D}
\begin{aligned}
   U(\vr_0,\vmu)&=\frac{\mu_0}{8\pi}
   \sum_{\bm n\neq\bm 0}
   \frac{\vmu\cdot\vmu_{\bm n}-3(\vmu\cdot\Rhat_{\bm n})(\vmu_{\bm n}\cdot\Rhat_{\bm n})}
        {|\vR_{\bm n}|^{3}},
\end{aligned}
\end{equation}
where
\begin{equation}
   \vR_{\bm n}=\vr_0-\vr_{\bm n},
   \qquad \Rhat_{\bm n}=\frac{\vR_{\bm n}}{|\vR_{\bm n}|}.
\end{equation}
Equation \eqref{eq:U3D} is the closed-form trap potential for arbitrary position and orientation. 
\begin{figure}[t!]
  \centering
  \includegraphics[width=\columnwidth]{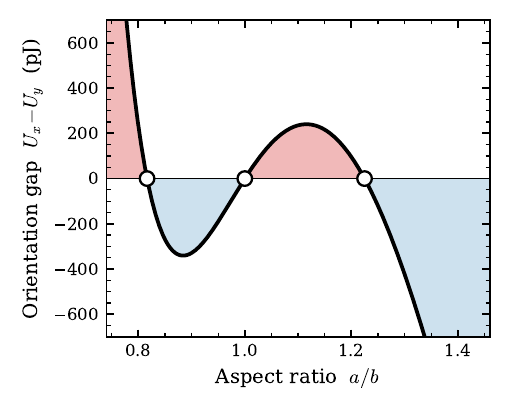}
  \caption{Orientational flip in the rectangular tube. The in-plane energy gap $U_x-U_y$ as a function of the aspect ratio $a/b$, with $a\ge b$ such that $\xhat$ is the long axis. The short axis $\yhat$ is the easy axis in the shaded interval $1<a/b<1.2249$ (positive gap); above $a/b\approx1.2249$ the long axis $\xhat$ is selected. By the $a\leftrightarrow b$ relabelling the same short-axis preference recurs in the interval $1/1.2249<a/b<1$, where $\xhat$ is now the shorter axis: this range is the geometric inverse $a/b\to(a/b)^{-1}$.}
  \label{fig:flip}
\end{figure}

\paragraph{Orientation as a diagonal quadratic form.}
For an exactly centred dipole, $\vr_0=\bm0$, the images lie at $\vr_{\bm n}=(2n_xa,2n_yb,2n_zc)$. Each cross term $\mu_\alpha\mu_\beta$ with $\alpha\neq\beta$ in \eqref{eq:U3D} is odd under reflection of either index it contains, $n_\alpha\mapsto-n_\alpha$, and cancels in pairs under the sum; this pairing preserves $\rho_{\bm n}$, so it is unaffected by the conditional convergence discussed below. The orientational energy is therefore diagonal in the trap axes (Appendix~A),
\begin{equation}
\label{eq:Uorient}
   U(\theta,\phi)=U_x\sin^2\!\theta\cos^2\!\phi
                 +U_y\sin^2\!\theta\sin^2\!\phi
                 +U_z\cos^2\!\theta,
\end{equation}
where $(\theta,\phi)$ are the polar and azimuthal angles of $\vmu$. The three coefficients are the principal self-energies
\begin{equation}
\label{eq:Ualpha}
   U_\alpha=\frac{\mu_0\mu^2}{64\pi}\!\!\sum_{\bm n\neq\bm 0}\!\!\,\,
   (-1)^{n_\alpha}\,
   \frac{\xi_\beta^2+\xi_\gamma^2-2\xi_\alpha^2}{\rho_{\bm n}^{5}},
\end{equation}
with $\xi_x=n_xa$, $\xi_y=n_yb$, $\xi_z=n_zc$, $\rho_{\bm n}=(\xi_x^2+\xi_y^2+\xi_z^2)^{1/2}$ (such that $|\vR_{\bm n}|=2\rho_{\bm n}$ at the centred source), $(\alpha\beta\gamma)$ a permutation of $(xyz)$, and $n_\alpha$ the index along axis $\alpha$. It follows that the easy axis is the trap axis of least $U_\alpha$, and that the librational stiffness about a minimum $\alpha$ toward a neighbour $\beta$ is $2(U_\beta-U_\alpha)$. The orientational behaviour is therefore fixed entirely by the ordering of $U_x$, $U_y$, $U_z$.

\paragraph{Regularisation.}
The summand of \eqref{eq:Ualpha} decreases as $\rho^{-3}$, the same borderline rate as the field sum above: it is conditionally convergent, but the alternating moments carry no net magnetisation, so its value is shape independent and equals the symmetric limit defined there.

\section{Short-axis alignment and the orientational flip}

In the limit $c\to\infty$ the trap becomes a tube of rectangular cross-section, infinite along $z$. The images with $n_z\neq0$ recede to infinity and the lattice \eqref{eq:lattice} collapses to its $n_z=0$ plane. For a centred dipole the in-plane energy is $U(\phi)=U_x\cos^2\phi+U_y\sin^2\phi$, and the in-plane librational stiffness is $2|U_x-U_y|$. The easy axis is set by the sign of the gap $U_x-U_y$.

The gap is a difference of two-dimensional lattice sums, and it separates into two physically distinct contributions. The images that lie on a coordinate axis of the cross-section ($n_x=0$ or $n_y=0$) contribute, in closed form,
\begin{equation}
\label{eq:gap_axis}
   (U_x-U_y)_{\rm axis}
   =\frac{\mu_0\mu^2\zeta(3)}{64\pi}\left(\frac1{a^3}-\frac1{b^3}\right),
\end{equation}
which is negative for $a>b$ and therefore favours alignment with the long axis. The images that lie off both axes contribute with the opposite sign; the nearest such images, at indices $(\pm1,\pm1)$, give
\begin{equation}
\label{eq:gap_diag}
   (U_x-U_y)_{\rm diag}^{(1,1)}
   =\frac{\mu_0\mu^2}{64\pi}\,\frac{12\,(a^2-b^2)}{(a^2+b^2)^{5/2}},
\end{equation}
which is positive for $a>b$ and favours the short axis. The easy axis is selected by the competition between these two contributions. Just above the square cross-section the off-axis sum dominates and the short axis is preferred; as $a/b$ increases the axis term \eqref{eq:gap_axis} overtakes it and the easy axis becomes the long axis. The change of sign occurs at $a/b\approx1.2249$ (equivalently $(a/b)\approx0.8164$), numerically close to, but distinct from, $\sqrt{3/2}=1.2247$. The threshold is a root of the full lattice sum, for which no closed form is known. Fig.~\ref{fig:flip} shows the gap and its change of sign. 

\section{Orientational phase diagram}

\begin{figure}[tb]
  \centering
  \includegraphics[width=\columnwidth]{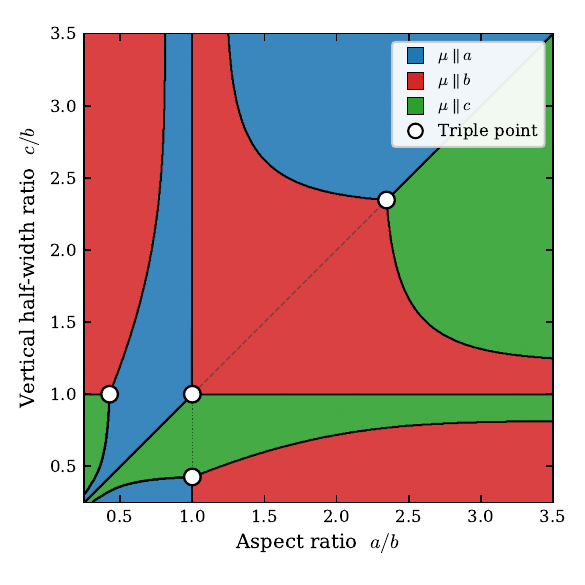}
  \caption{Equilibrium orientation of a centred dipole in the closed trap, as a function of the two half-width ratios $a/b$ and $c/b$ ($a$, $b$, $c$ the half-widths along $x$, $y$, $z$). Colour indicates the easy axis, that is whether the moment aligns with the $a$-, $b$- or $c$-edge of the trap. The domains meet at the four triple points $U_x=U_y=U_z$ (circles): the cube $(1,1)$ and the symmetry orbit of the accidental point on the diagonal at $\lambda_c=2.346182$. As $c/b$ increases the trap approaches the tube limit.}
  \label{fig:phase}
\end{figure}

In the closed trap all three coefficients in \eqref{eq:Uorient} compete, and the equilibrium orientation depends on the two half-width ratios $a/b$ and $c/b$. Fig.~\ref{fig:phase} shows the easy axis over this plane. The orientation becomes degenerate at triple points, $U_x=U_y=U_z$, which can be enumerated: each mirror plane $a=b$, $b=c$, $a=c$ is an exact component of the corresponding degeneracy locus, reducing the search to one-dimensional conditions whose roots are then certified numerically. The result is four triple points, the cube $a=b=c$ (fixed by octahedral symmetry) and one symmetry orbit of accidental points of shape $1\!:\!\lambda_c\!:\!\lambda_c$; no degeneracy with three distinct edge lengths occurs. The three members of this orbit are the same trap shape with the short edge assigned to each axis in turn, related by axis relabelling and the inversion $\lambda\to\lambda^{-1}$; they are geometrically equivalent rather than physically distinct, exactly as the two short-axis intervals of Fig.~\ref{fig:flip} are inverse images of one another. The accidental ratio $\lambda_c=2.346182$ is protected by no point-group symmetry, and no closed algebraic form is known: in particular $\lambda_c^{2}=5.5046\neq 11/2$, so $\sqrt{11/2}$ is a truncation coincidence. The phase diagram also identifies the regime in which the two-dimensional description applies. As $c/b$ increases, the coefficients $U_x$ and $U_y$ approach their tube values, and the easy axis is governed by the in-plane competition of Fig.~\ref{fig:flip}.

\section{Finite-element verification}
\label{sec:fem}

\begin{figure}[t]
  \centering
  \includegraphics[width=\columnwidth]{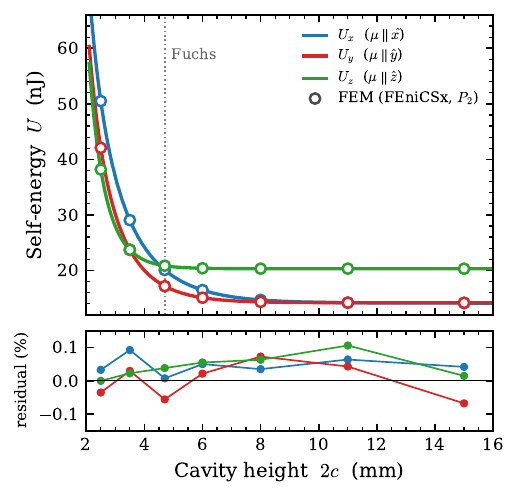}
  \caption{Comparison of the closed form with a finite-element solution.
  The principal self-energies $U_x$, $U_y$ and $U_z$ are computed at the
  cross-section $a=2.25$, $b=1.75$~mm as the full height $2c$ varies.
  Solid curves: the image-lattice result \eqref{eq:U3D}. Open circles: a
  finite-element solution of the Neumann problem in FEniCSx with quadratic
  elements~\cite{FEniCSx2023}. The residuals (lower panel) remain below
  $0.16\%$. The energies flatten to the tube limit at large $2c$, and the
  crossing of $U_x$ and $U_y$ is the flip of Fig.~\ref{fig:flip}.}
  \label{fig:fem}
\end{figure}

Equation \eqref{eq:U3D} is an exact statement about the Meissner boundary-value problem. We verify it by solving the same Neumann problem directly with finite elements, using two independent implementations: a FEniCSx discretisation~\cite{FEniCSx2023} and a scikit-fem solver, \textsc{sctrap}~\cite{scikitfem2020,sctrap}, which agree with each other and with the closed form. Fig.~\ref{fig:fem} compares the analytic and finite-element results along a sweep of the trap height at fixed cross-section. The agreement is better than $0.16\%$ at every point, with a residual set by the finite mesh resolution. Being a direct solution of the boundary-value problem, the finite-element result involves no lattice sum, and so confirms the regularised value independently of any summation order.

\section{Parallel-plate limit}

One would expect that sending two pairs of the side walls to infinity recovers the parallel-plate trap of Ref.~\cite{HeadleyMagnet}. Indeed, sending $a,b\to\infty$ at fixed $c$ pushes every image with $n_x\neq0$ or $n_y\neq0$ to infinite distance and leaves the single $z$-axis chain $n_x=n_y=0$. The surviving sums are one-dimensional Dirichlet series, and the transverse and normal coefficients reduce to
\begin{equation}
\label{eq:plate_coeffs}
   U_x=U_y\to\frac{\mu_0\mu^2}{64\pi c^3}\,2\zeta(3),
   \qquad
   U_z\to\frac{\mu_0\mu^2}{64\pi c^3}\,3\zeta(3).
\end{equation}
Because $U_x=U_y$, the orientational energy \eqref{eq:Uorient} depends only on the tilt $\beta_0$ out of the plates, $U(\beta_0)\propto2+\sin^2\!\beta_0$, with its minimum at $\beta_0=0$: the moment lies parallel to the plates. This is exactly the centred-dipole potential of Ref.~\cite{HeadleyMagnet} at plate separation $2c$; the reduction, including the prefactor, is carried out in Appendix~B.

\section{Conclusion}

The image method reduces the magnetic trapping problem in a closed superconducting trap to a single lattice sum, \eqref{eq:U3D}, valid on all six walls. The orientational energy is a diagonal quadratic form whose coefficients are Epstein-zeta-type sums, \eqref{eq:Uorient} and \eqref{eq:Ualpha}, and the easy-axis selection is non-monotonic in the trap shape: a centred dipole aligns with the short cross-sectional axis over a finite range of aspect ratios before the long axis is selected. In the closed trap this behaviour is organised by a phase diagram with exactly four triple points, the cube and one symmetry orbit of accidental points whose ratio $\lambda_c=2.346182$ is a zero of a difference of Epstein zeta functions with no known closed algebraic form. The closed form supplies the full translational and orientational stiffness, and hence all trap and librational frequencies. It agrees with a finite-element solution of the Neumann problem to better than $0.16\%$.

\section{Acknowledgements} We thank Tim Fuchs, Hendrik Ulbricht, Fabian Müller, and Dorian Amaral for helpful discussions, and Enrique Verduras for insight into Epstein zeta functions. We also thank Sophie Heinrich for feedback on the draft. We acknowledge the EU EIC Pathfinder project QuCoM (101046973).

\section{Data availability} The image-method and finite-element code, the reference data, and the scripts that regenerate every figure are openly available~\cite{cuboidalcavitytrap}.

\bibliographystyle{apsrev4-1}
\bibliography{refs_3d}

\section{Appendix A: Diagonalisation of the centred orientational energy}
\label{app:diag}

For a centred source the trap potential \eqref{eq:U3D} is a quadratic form in the moment,
\begin{equation}
\label{eq:quadform}
   U(\vmu)=\sum_\alpha U_{\alpha\alpha}\,\mu_\alpha^2
          +\sum_{\alpha<\beta}U_{\alpha\beta}\,\mu_\alpha\mu_\beta .
\end{equation}
We show that the off-diagonal coefficients $U_{\alpha\beta}$ vanish, so that the form is diagonal in the trap axes with $U_\alpha=\mu^2U_{\alpha\alpha}$ the principal self-energies of \eqref{eq:Ualpha}.

At $\vr_0=\bm0$ the images lie at $\vR_{\bm n}=-2(\xi_x,\xi_y,\xi_z)$ in the notation of \eqref{eq:Ualpha}, so $\Rhat_{\bm n}=-(\xi_x,\xi_y,\xi_z)/\rho_{\bm n}$ and $|\vR_{\bm n}|=2\rho_{\bm n}$. The unit vector $\Rhat_{\bm n}$ enters \eqref{eq:U3D} only through the symmetric product $(\vmu\cdot\Rhat_{\bm n})(\vmu_{\bm n}\cdot\Rhat_{\bm n})$, so its sign does not matter. Here, Latin indices $i,j$ run as summation dummies over the Cartesian components $\{x,y,z\}$, while the Greek $\alpha,\beta,\gamma$ denote a fixed permutation of the axes, as in \eqref{eq:Ualpha}. Expanding the two dot products,
\begin{align}
\label{eq:dots}
   \vmu\cdot\vmu_{\bm n}&=\sum_i(-1)^{n_i}\mu_i^2,\\
   (\vmu\cdot\Rhat_{\bm n})(\vmu_{\bm n}\cdot\Rhat_{\bm n})
   &=\frac{1}{\rho_{\bm n}^2}\sum_{i,j}(-1)^{n_j}\,\xi_i\xi_j\,\mu_i\mu_j .
\end{align}

\paragraph{Off-diagonal terms.}
The cross monomial $\mu_\alpha\mu_\beta$ ($\alpha\neq\beta$) comes only from the second term of the numerator, through $(i,j)=(\alpha,\beta)$ and $(\beta,\alpha)$. With $|\vR_{\bm n}|^{3}\rho_{\bm n}^{2}=8\rho_{\bm n}^{5}$,
\begin{equation}
\label{eq:offdiag}
   U_{\alpha\beta}
   =-\frac{3\mu_0}{64\pi}\sum_{\bm n\neq\bm 0}
     \frac{\xi_\alpha\xi_\beta\,[(-1)^{n_\alpha}+(-1)^{n_\beta}]}{\rho_{\bm n}^{5}} .
\end{equation}
Under reflection of the summation lattice in its $\alpha$-axis, $n_\alpha\mapsto-n_\alpha$, the factor $\xi_\alpha$ changes sign while $\xi_\beta$, the parities, and $\rho_{\bm n}$ are unchanged; the summand is therefore odd. Terms with $n_\alpha=0$ already vanish, and on the rest the reflection pairs each term with its negative. Since it preserves $\rho_{\bm n}$, the two members of a pair lie on the same shell, so the cancellation is exact within every symmetric truncation $\rho_{\bm n}\le\Lambda$ and does not rearrange the conditionally convergent sum. Hence $U_{\alpha\beta}=0$ and the form is diagonal.

\paragraph{Diagonal terms.}
The square $\mu_\alpha^2$ comes from $i=j=\alpha$ in \eqref{eq:dots}, with coefficient $(-1)^{n_\alpha}(1-3\xi_\alpha^2/\rho_{\bm n}^2)$. Using $\rho_{\bm n}^2-3\xi_\alpha^2=\xi_\beta^2+\xi_\gamma^2-2\xi_\alpha^2$ for $(\alpha\beta\gamma)$ a permutation of $(xyz)$,
\begin{equation}
\label{eq:diag}
   U_{\alpha\alpha}
   =\frac{\mu_0}{64\pi}\sum_{\bm n\neq\bm 0}(-1)^{n_\alpha}
     \frac{\xi_\beta^2+\xi_\gamma^2-2\xi_\alpha^2}{\rho_{\bm n}^{5}} ,
\end{equation}
so that $U_\alpha=\mu^2U_{\alpha\alpha}$ is Eq.~\eqref{eq:Ualpha}, and writing $\vmu=\mu\hat{\vmu}$ in polar angles gives the diagonal form \eqref{eq:Uorient}. A direct evaluation of \eqref{eq:U3D} on a non-cubic trap confirms $U_{\alpha\beta}=0$ to machine precision under symmetric truncation, with the cross terms reappearing for an off-centre source or an asymmetric summation order.

\section{Appendix B: Reduction to the parallel-plate limit}
\label{app:plate}

Sending both pairs of side walls to infinity, $a,b\to\infty$ at fixed $c$, recovers the parallel-plate result of Ref.~\cite{HeadleyMagnet} from the centred-trap coefficients \eqref{eq:Ualpha}. The images sit at $\vr_{\bm n}=(2n_xa,2n_yb,2n_zc)$, so every image with $n_x\neq0$ or $n_y\neq0$ has $\rho_{\bm n}\to\infty$, where the summand of \eqref{eq:Ualpha} falls as $\rho_{\bm n}^{-3}$ and vanishes. Only the $z$-axis chain $n_x=n_y=0$ survives, with $\xi_x=\xi_y=0$, $\xi_z=n_zc$ and $\rho_{\bm n}=|n_z|c$, reducing the lattice sums to Dirichlet series. For the coefficient along the chain,
\begin{equation}
   U_z=\frac{\mu_0\mu^2}{64\pi}\sum_{n_z\neq0}(-1)^{n_z}\frac{-2}{|n_z|^{3}c^{3}}
      =-\frac{\mu_0\mu^2}{32\pi c^3}\sum_{n_z\neq0}\frac{(-1)^{n_z}}{|n_z|^{3}},
\end{equation}
and with $\sum_{n_z\neq0}(-1)^{n_z}|n_z|^{-3}=-2\,\eta(3)=-\tfrac32\zeta(3)$, where $\eta(s)=(1-2^{1-s})\zeta(s)$ is the Dirichlet eta function,
\begin{equation}
   U_z\;\longrightarrow\;\frac{\mu_0\mu^2}{64\pi c^3}\,3\zeta(3).
\end{equation}
For the two transverse coefficients $(-1)^{n_x}=1$ and the numerator of \eqref{eq:Ualpha} is $\xi_z^2$, so
\begin{equation}
   U_x=U_y=\frac{\mu_0\mu^2}{64\pi}\sum_{n_z\neq0}\frac{1}{|n_z|^{3}c^{3}}
          \;\longrightarrow\;\frac{\mu_0\mu^2}{64\pi c^3}\,2\zeta(3).
\end{equation}
With $U_x=U_y$ the orientational energy \eqref{eq:Uorient} depends only on the tilt out of the plates. Writing $\beta_0=\tfrac{\pi}{2}-\theta$ for the angle measured from the plate plane, so that $\sin\theta=\cos\beta_0$ and $\cos\theta=\sin\beta_0$,
\begin{equation}
\label{eq:plate}
   U(\beta_0)=U_x\cos^2\!\beta_0+U_z\sin^2\!\beta_0
           =\frac{\mu_0\mu^2\zeta(3)}{64\pi c^3}\bigl(2+\sin^2\!\beta_0\bigr).
\end{equation}
Identifying the half-width $c$ with the plate half-separation, so that the plate spacing is $2c$, Eq.~\eqref{eq:plate} is exactly the centred-dipole potential of Ref.~\cite{HeadleyMagnet}, $U(\beta_0)=\tfrac{\mu_0\mu^2\zeta(3)}{128\pi c^3}(5-\cos2\beta_0)$, on using $5-\cos2\beta_0=4+2\sin^2\!\beta_0$. The energy is minimised at $\beta_0=0$, the moment lying parallel to the plates, with librational stiffness $2(U_z-U_x)=\mu_0\mu^2\zeta(3)/32\pi c^3$.

\end{document}